\documentstyle[12pt,epsfig]{article}

\begin{document}
%
%
\newcommand{\beq}{\begin{equation}}
\newcommand{\eeq}{\end{equation}}
\newcommand{\beqn}{\begin{eqnarray}}
\newcommand{\eeqn}{\end{eqnarray}}
\def\sss{\scriptscriptstyle}
\newcommand{\mw}{M_{\sss W}}
\newcommand{\mz}{M_{\sss Z}}
\def\ie{{\it i.e.\/}}
\def\eg{{\it e.g.\/}}
\def\rffly#1{\mathrel{\raise.3ex\hbox{$#1$\kern-.75em\lower1ex\hbox{$\sim$}}}}
\def\lsim{\rffly<}
\def\gsim{\rffly>}
%
%
\def\ijmp#1#2#3{{\it Int.\ J.\ Mod.\ Phys.} {\bf A#1} (19#2) #3}
\def\mpla#1#2#3{{\it Mod.\ Phys.\ Lett.} {\bf A#1}, (19#2) #3}
\def\npb#1#2#3{{\it Nucl.\ Phys.} {\bf B#1} (19#2) #3}
\def\plb#1#2#3{{\it Phys.\ Lett.} {\bf #1B} (19#2) #3}
\def\prd#1#2#3{{\it Phys.\ Rev.} {\bf D#1} (19#2) #3}
\def\prl#1#2#3{{\it Phys.\ Rev.\ Lett.} {\bf #1} (19#2) #3}
\def\zpc#1#2#3{{\it Zeit.\ Phys.} {\bf C#1} (19#2) #3}
\def\arnps#1#2#3{{\it Ann.\ Rev.\ Nucl.\ Part.\ Sci.} {\bf #1}, (19#2) #3}
\def\nci#1#2#3{{\it Nuovo Cimento} {\bf #1} (19#2) #3}
%
%
\def\dbeta{\beta\beta_{0\nu}}
\def\eeWW{e^-e^-\to W^-W^-}
\def\llWW{\ell^-\ell^-\to W^-W^-}
\def\ggllww{\gamma\gamma \to \ell^+ \ell^+ W^- W^-}
\def\ggmmww{\gamma\gamma \to \mu^+ \mu^+ W^- W^-}
\def\ggeeww{\gamma\gamma \to e^+ e^+  W^- W^-}
\def\ggttww{\gamma\gamma \to \tau^+ \tau^+ W^- W^-}
\def\egnllw{e^- \gamma \to \nu_e \ell^- \ell^- W^+}
\def\egnmmw{e^- \gamma \to \nu_e \mu^- \mu^- W^+}
\def\egnttw{e^- \gamma \to \nu_e \tau^- \tau^- W^+}
\def\eennll{e^- e^- \to \nu_e \nu_e \ell^- \ell^-}
\def\egeww{e^- \gamma \to e^+ W^- W^-}
\def\eeeeww{e^+ e^- \to e^+ e^+ W^- W^-}
\def\ggeeww{\gamma\gamma \to e^+ e^+ W^- W^-}
\begin{titlepage}
\begin{center}
{ \Large \bf Inverse Neutrinoless Double Beta Decay Revisited}
\\
\vspace*{1.cm}
\begin{tabular}[t]{c}
{\bf G. B\'elanger$^a$, F. Boudjema$^a$, D. London$^b$ and H. Nadeau$^c$}\\
\\
$^a${\it Laboratoire de Physique Th\'eorique}
EN{\large S}{\Large L}{\large A}PP
\footnote{URA 14-36 du CNRS, associ\'ee \`a l'E.N.S de Lyon et \`a
l'Universit\'e de Savoie.}\\
\smallskip
{\it Chemin de Bellevue, B.P. 110, F-74941 Annecy-le-Vieux, Cedex, France.}
\\
$^b${\it Laboratoire de Physique Nucl\'eaire, Universit\'e
de Montr\'eal, } \\
\smallskip
{\it C.P. 6128, Montr\'eal, Qu\'ebec, Canada, H3C 3J7.}
\\
$^c${\it Physics Department, McGill University, }\\
{\it 3600 University St., Montr\'eal, Qu\'ebec, Canada, H3A 2T8.}
\\
\end{tabular}
\end{center}
\vspace*{\fill}

\centerline{ {\bf Abstract} }
\baselineskip=14pt
\noindent
{\small

We critically reexamine the prospects for the observation of the $\Delta
L=2$ lepton-number-violating process $\eeWW$ using the $e^-e^-$ option of a
high-energy $e^+e^-$ collider (NLC). We find that, except in the most
contrived scenarios, constraints from neutrinoless double beta decay render
the process unobservable at an NLC of $\sqrt{s}<2$ TeV. Other $\Delta L=2$
processes such as $\ggllww$, $\egnllw$, $\eennll$ ($\ell=\mu,\tau$), and
$\egeww$, which use various options of the NLC, require a $\sqrt{s}$ of at
least 4 TeV for observability.

}
\vspace*{\fill}

\vspace*{0.1cm}
\rightline{ENSLAPP-A-537/95}
\rightline{UdeM-GPP-TH-95-30}
\rightline{\hfil McGill/95-37}
\rightline{hep-ph/959508317}
\rightline{August 1995}
\vspace*{\fill}
\noindent
{\footnotesize $\;^{*}$ URA 14-36 du CNRS, associ\'ee \`a l'E.N.S de
Lyon et \`a l'Universit\'e de Savoie.}
\end{titlepage}
\baselineskip=18pt
\setcounter{section}{1}
\setcounter{subsection}{0}
\setcounter{equation}{0}
\def\thesubsection {\thesection.\arabic{subsection}}
\def\theequation{\thesection.\arabic{equation}}
\setcounter{equation}{0}
\def\thequation{\thesection.\arabic{equation}}
\setcounter{section}{0}
\setcounter{subsection}{0}

\section{Introduction}

One of the most intriguing puzzles in modern particle physics is whether
the neutrino has a mass. In fact, it is doubly interesting since, if the
neutrino is massive, one will want to know whether it has a Dirac or a
Majorana mass. If the neutrino has a Majorana mass, then it will contribute
to $\Delta L=2$ lepton-number-violating processes such as neutrinoless
double beta decay ($\dbeta$). The key subprocess in $\dbeta$ is $W^- W^-
\to e^- e^-$, mediated by a Majorana $\nu_e$.

One possible future collider which is being vigorously investigated at the
moment is a high-energy linear $e^+e^-$ collider, known generically as the
Next Linear Collider (NLC). With such a collider, it is possible to replace
the positron by another electron and look at $e^-e^-$ collisions. If the
electron neutrino has a Majorana mass, it may be possible to observe the
process $\eeWW$. This is essentially the inverse of neutrinoless double
beta decay.

In fact, this is not a new idea. The process $\eeWW$ has been looked at
several times, by different authors, over the last decade or so
\cite{Rizzo}-\cite{HM}. In the most recent analysis, the authors of
Ref.~\cite{HM} found that this process could be observable at an NLC of
$\sqrt{s}=500$ GeV or 1 TeV. One of the purposes of the present paper is to
reexamine this analysis. Once the constraints from $\dbeta$ are taken into
account, we find that, in fact, except for extremely contrived scenarios,
the cross section for $\eeWW$ is simply too small for it to be seen at a
500 GeV or 1 TeV NLC. An NLC of at least $\sqrt{s}=2$ TeV will be necessary
in order to have a hope of observing this process.

The limits from $\dbeta$ apply only to $\nu_e$. Should the $\nu_\mu$ have a
Majorana mass, it will contribute to the processes $\mu^-\mu^- \to W^-W^-$
and its inverse (and similarly for the $\nu_\tau$), with no constraints
from low-energy processes. However, unless a $\mu^-\mu^-$ collider is built
\cite{mumu}, such lepton-number-violating processes cannot take place
directly. Fortunately, there are other possibilities at the NLC. It is
possible to backscatter laser light off one or both of the beams, creating
an $e\gamma$ or $\gamma\gamma$ collider \cite{backscatter}. $\mu^-\mu^- \to
W^-W^-$ can then be observed as a subprocess in one of the various modes of
the NLC. For example, the observation of $\gamma\gamma \to \mu^+\mu^+
W^-W^-$ would be evidence for a Majorana $\nu_\mu$. This is the second
purpose of the paper -- to investigate the possibilities for the detection
of $\Delta L=2$ lepton-number violation in the muon or tau sectors at the
NLC. We will see that an NLC with a centre-of-mass energy of at least 4 TeV
is necessary.

The paper is organized as follows. In the following section we discuss the
process $\eeWW$, paying careful attention to the constraints from unitarity
and $\dbeta$. In Sec.~3 we elaborate on the possibilities for detecting
$\Delta L=2$ lepton-number violation in the muon or tau sectors. Sec.~4
contains a discussion of the prospects for detecting a Majorana $\nu_e$ if
no $e^-e^-$ collider is ever built. We conclude in Sec.~5.

\section{$\eeWW$}

\subsection{Neutrino Mixing}

Suppose that the $\nu_e$ mixes with other neutrinos. For the moment, we
leave the number of new neutrinos unspecified, as well as their
transformation properties under $SU(2)_{\sss L}$. (The $\nu_e$ could even
mix with $\nu_\mu$ and/or $\nu_\tau$, although this will lead to
flavour-changing neutral currents, which are extremely stringently
constrained.) Once the mass matrix is diagonalized, $\nu_e$ can be
expressed in terms of the mass eigenstates $N_i$:
\beq
\nu_e = \sum_i U_{ei} N_i ~,
\eeq
where the mixing matrix $U$ is unitary. Phenomenologically, we have
observed two things. First, the $\nu_e$ does not mix much with other
neutrinos \cite{Nardi}:
\beq
\sum_{i\ne 1} |U_{ei}|^2 < 6.6 \times 10^{-3} ~~(90\%~c.l.)~.
\label{mixinglimit}
\eeq
This limit is essentially independent of the $SU(2)_{\sss L}$
transformation properties of the neutrino(s) with which the $\nu_e$ mixes.
Also, the limit is quite conservative -- it allows for the possibility that
the other charged fermions also mix with new, exotic charged particles
\cite{LL}. If one assumes that the only new particles are neutrinos, then
the above limit improves somewhat to $5.0 \times 10^{-3}$. Thus, the
$\nu_e$ is mainly $N_1$. Second, from muon decay, we know that the $N_1$ is
very light: $M_1 < 7$ eV \cite{PDG}.


\subsection{Cross Section for $\eeWW$}

\begin{figure*}[thb]
 \hspace*{-2.1cm}
 \vspace*{.3cm}
 \mbox{\epsfxsize=18cm\epsfysize=20.cm\epsffile{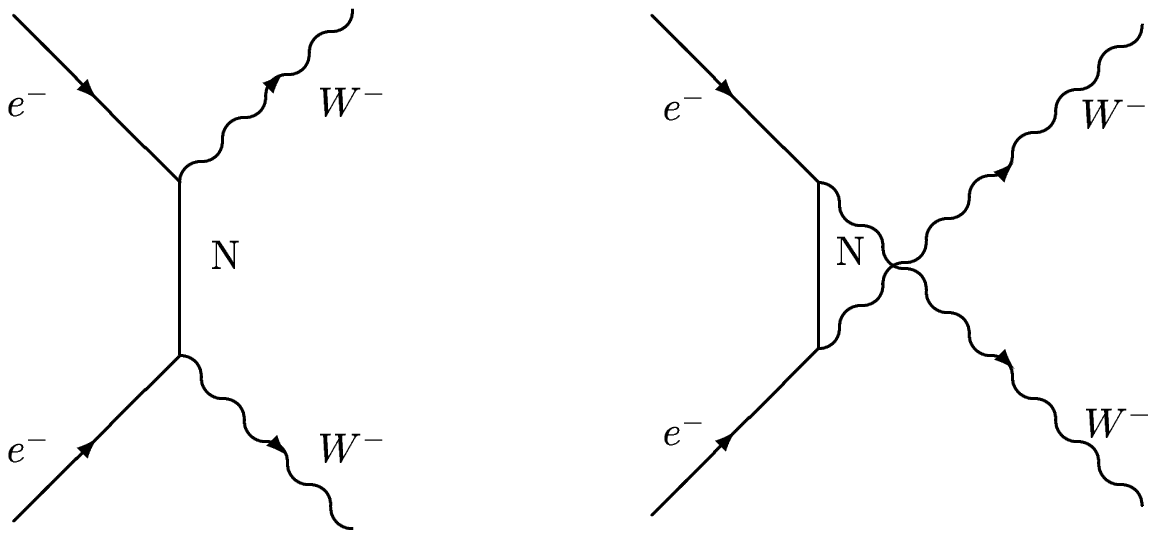}}
 \vspace*{-15.7cm}
\caption{\label{eeWWfigs}{\em Diagrams contributing to $\eeWW$.}}
\end{figure*}

Assuming that the $N_i$ are Majorana neutrinos, they will contribute
to the process $\eeWW$ through the diagrams of Fig.~\ref{eeWWfigs}.
(If right-handed $W$'s exist, they can also be produced, either singly
or in pairs, through similar diagrams. In this paper we consider only
ordinary $W$'s in the final state -- the production of $W_{\sss R}$'s
is discussed in Refs.~\cite{Rizzo,MPV,Rizzo2}.) Neglecting the
electron mass, the differential cross section for unpolarized
electrons is
\newpage
\beqn
{d\sigma \over d\cos\theta} & = & {g^4\over 512\pi s}
\left( 1 - {4\mw^2\over s}\right)^{1/2}
\sum_{ij} M_i M_j \left(U_{ei}\right)^2 \left(U_{ej}\right)^2 \nonumber \\
& & ~~~~\times \left[ {1\over (t-M_i^2)(t-M_j^2)}
\left( {(s-2\mw^2)(t-\mw^2)^2 \over 2\mw^4} + {(t-\mw^2)(u-\mw^2)\over \mw^2}
+{s\over 2} \right) \right. \nonumber \\
& & ~~~~~~~~+ ~u\leftrightarrow t
+ \left( {1\over (t-M_i^2)(u-M_j^2)} +
{1\over (t-M_j^2)(u-M_i^2)} \right) \nonumber \\
& & ~~~~~~~~~~~
\times \left.
\left( {(tu-\mw^4)(s-2\mw^2) \over 2\mw^4} - {(t-\mw^2)(u-\mw^2)\over \mw^2}
+{3s\over 2} \right) \right].
\label{Xsection}
\eeqn
Although this expression is rather complicated, it simplifies
considerably in the limit that $s \gg \mw^2$, which is a reasonable
approximation for the NLC. In this case, the terms in the square
brackets which are proportional to $1/\mw^4$ dominate -- they are
larger than the terms proportional to $1/\mw^2$ by a factor $\sim
s/\mw^2$. Keeping only the dominant terms, the differential cross
section then becomes simply
\beq
{d\sigma \over d\cos\theta} = {g^4\over 1024\pi\mw^4}
\left( \sum_i M_i \left(U_{ei}\right)^2
\left[ {t\over(t-M_i^2)} + {u\over(u-M_i^2)} \right] \right)^2 ~.
\label{Xsectionlong}
\eeq
(Although this is an excellent approximation to the differential cross
section, we nevertheless use the full expression (Eq.~\ref{Xsection})
when presenting numerical results.)

Note that Eq.~\ref{Xsectionlong} is precisely what is obtained if one
calculates the cross section for the process $e^- e^- \to W^-_{long}
W^-_{long}$, where $W^-_{long}$ is the longitudinal component of the
$W^-$, as we have verified by explicit calculation. We thus confirm
the observation of Ref.~\cite{HM} that the production of longitudinal
$W$'s dominates the process $\eeWW$ if $s\gg\mw^2$. The full helicity
amplitudes are given in the Appendix.

There are two limiting cases of Eq.~\ref{Xsectionlong} which will be
useful in what follows. First, if $s \gg M_i^2$, the cross
section becomes
\beq
{d\sigma \over d\cos\theta} = {g^4\over 256\pi\mw^4}
\left( \sum_i M_i \left(U_{ei}\right)^2 \right)^2 ~.
\label{Xsectionunitarity}
\eeq
Second, in the other limit, $M_i^2 \gg s$, we get
\beq
{d\sigma \over d\cos\theta} = {g^4\over 1024\pi\mw^4} s^2
\left( \sum_i {\left(U_{ei}\right)^2 \over M_i } \right)^2 ~.
\label{XsectionheavyN}
\eeq
Note that, in this limit, the cross section grows like $s^2$, as was
observed in Ref.~\cite{HM}.

\subsection{Unitarity Considerations}

{}From Eq.~\ref{Xsectionunitarity}, we see that, in the high-energy limit ($s
\to \infty$), the cross section tends towards a constant:
\beq
\sigma_{s\to\infty} = {g^4\over 128\pi\mw^4}
\left( \sum_i M_i \left(U_{ei}\right)^2 \right)^2 ~.
\label{unitarity}
\eeq
In this particular case this indicates a violation of unitarity, since
the amplitude (which is a pure s-wave) grows as $\sqrt{s}$.

There are basically two ways in which this unitarity violation can be
cured. The first through the inclusion of a Higgs triplet. If the
neutrinos with which the $\nu_e$ mixes are $SU(2)_{\sss L}$ doublets,
then they can acquire Majorana masses by giving the Higgs triplet a
vacuum expectation value (v.e.v.). This Higgs triplet includes a
doubly-charged Higgs, $H^{--}$. In this case unitarity is restored
through the inclusion of a diagram in which the $H^{--}$ is exchanged
in the s-channel.

However, this type of solution has been virtually eliminated
phenomenologically. The v.e.v.\ of the Higgs triplet breaks lepton number
spontaneously, producing a Majoron. But light Majorons would contribute
significantly in $Z$ decays, and have been ruled out by the precision LEP
data. Such models are therefore untenable. There are ways to evade the LEP
bounds -- for instance, one can add a Higgs singlet and allow the triplet
to mix with the singlet \cite{ChoiSan}. However, in addition to being
somewhat artificial, this solution does not explain the large range of
neutrinos masses. If all neutrinos are $SU(2)_{\sss L}$ doublets, then all
their masses would be Majorana, and would come from the v.e.v.\ of the
Higgs triplet. Precision measurements on the $Z$ peak constrain such a
v.e.v.\ to be at most a few percent of that of the standard Higgs doublet
\cite{PDGLang}. It would therefore require an extremely large Yukawa
coupling to produce a neutrino mass in the TeV range. Such large Yukawa
couplings typically lead to other problems, such as the breakdown of
perturbation theory, etc. In addition, there is no natural explanation why
some neutrino masses should be in the eV range, while others are in the TeV
range. Not even the charged fermions of the standard model cover such a
large range in mass. For all of the above reasons, we discard the Higgs
triplet as a solution to the unitarity problem in $\eeWW$.

In the absence of Higgs triplets, the only way to restore unitarity is to
require that the neutrinos' masses and mixing angles satisfy
\beq
\sum_i \left(U_{ei}\right)^2 M_i = 0 ~.
\label{massmixrelation}
\eeq
Although this relation may appear arbitrary at first sight, it is in fact
automatically satisfied. It is straightforward to show that
\beq
\sum_i \left(U_{ei}\right)^2 M_i = M^*_{ee}~,
\eeq
where $M_{ee}$ is the Majorana mass of the $\nu_e$. However, because there
are no Higgs triplets, this mass is equal to zero, so that
Eq.~\ref{massmixrelation} holds.

As an explicit example, consider the famous seesaw mechanism: one adds a
right-handed neutrino $N_{\sss R}$ to the spectrum. This neutrino acquires
a large Majorana mass $M$ through the v.e.v.\ of a Higgs singlet, and the
combination ${\bar N}_{\sss R}\nu_{e\sss L} + h.c.$ obtains a Dirac mass
$m$ once the ordinary Higgs doublet gets a v.e.v. The mass matrix looks
like
\beq
\left(\matrix{0 & m \cr m & M \cr}\right).
\eeq
The two mass eigenstates are $N_1$ and $N_2$, with masses $-m^2/M$ and $M$,
respectively (the minus sign in front of $M_1$ can be removed by a
$\gamma_5$ rotation). For $m$ of the order of the electron mass and $M$
about 1 TeV, one obtains a mass of about 1 eV for the lightest neutrino.
(Thus, in such models, the large range of neutrino masses is explained in a
natural way, unlike the Higgs triplet models.) The $\nu_e$ is a linear
combination of these two physical neutrinos:
\beq
\nu_e = \cos\theta N_1 + \sin\theta N_2~,~~~~~\sin\theta = {m\over M}~.
\eeq
It is clear that, with these masses and mixing angles, the relation in
Eq.~\ref{massmixrelation} is automatically satisfied.

The downside of this particular solution is that the mixing of the
$\nu_e$ with the $N$ is tiny: for $m \sim m_e$ and $M \sim 1$ TeV,
$\sin\theta \sim 10^{-6}$! This would make the cross section for
$\eeWW$ invisible, since $\sigma(\eeWW) \propto \sin^4 \theta$, and is
typical of what happens in left-right symmetric models \cite{Rizzo}.
However, if the $\nu_e$ mixes with more neutrinos, it is, in
principle, possible to satisfy Eq.~\ref{massmixrelation} without
having such small values of $U_{ei}$. (This is the assumed solution in
Ref.~\cite{DKR}.) This is perhaps a bit artificial, and probably
requires some fine-tuning, but it is possible. If this is how
unitarity restoration comes about, then Eq.~\ref{Xsection} contains
all the contributions to $\eeWW$.

Although it is interesting to understand how unitarity is restored in
different models, the above discussion demonstrates that the cross
section for $\eeWW$ is essentially unconstrained by such
considerations -- the $U_{ei}$ and $M_i$ can take any values
consistent with the phenomenological limits in Sec.\ 2.1. This is not
the case when the experimental limits on neutrinoless double beta
decay are taken into account, which we do in the next subsection.

\subsection{Limits from $\dbeta$}

As mentioned in the introduction, $\eeWW$ is essentially the inverse
of neutrinoless double beta decay. We might therefore expect that the
limits on the latter process could constrain the former.

If some of the neutrinos have masses $M_i \ll 1$ GeV, then, for these
neutrinos, the quantity which contributes to $\dbeta$ is
\beq
\langle m_\nu \rangle = \sum_i{}' \left(U_{ei}\right)^2 M_i~,
\eeq
where the sum is over the light neutrinos. (For simplicity, we have
ignored factors corresponding to complications from the nuclear matrix
elements -- their inclusion does not change our conclusions. For more
details we refer the reader to Ref.~\cite{dbetarefs}.) The experimental
limit on $\langle m_\nu \rangle$ is \cite{dbetarefs}
\beq
\langle m_\nu \rangle \lsim 1~{\hbox{eV}}~.
\label{lightNlimit}
\eeq

As for the neutrinos which are heavy, $M_i \gg 1$ GeV, they can still
mediate $\dbeta$ decay. In this case the relevant quantity is
\beq
\langle m_\nu^{-1} \rangle_{\sss H} = \sum_i{}^{\prime\prime}
\left(U_{ei}\right)^2 {1\over M_i} ~,
\eeq
where the sum is over the heavy neutrinos. Now the experimental limit
on $\dbeta$ implies the following:
\beq
\langle q^2 \rangle
\sum_i{}^{\prime\prime} \left(U_{ei}\right)^2 {1\over M_i}
\lsim 1~{\hbox{eV}}~,
\eeq
where $q$ is an average nuclear momentum transfer. If one takes $q$
to be roughly about 100 MeV, one obtains the right order-of-magnitude
constraint. However, a more careful calculation, including all the
nuclear effects, gives \cite{Boris}
\beq
\sum_i{}^{\prime\prime} \left(U_{ei}\right)^2 {1\over M_i} <
5 \times 10^{-5}~{\hbox{TeV}}^{-1} ~.
\label{heavyNlimit}
\eeq
Assuming no cancellations, this implies a generic lower bound on the mass
of the heavy neutrino:
\beq
M_i > 2 \times 10^4 \left(U_{ei}\right)^2 ~{\hbox{TeV}}~.
\label{lowerNlimit}
\eeq
For $\left(U_{ei}\right)^2 \sim 5 \times 10^{-3}$, this gives $M_i > 100$
TeV!

However, it is possible to evade this order-of-magnitude bound if one
allows cancellations among the various terms. This can come about in one of
two ways: either (i) all the heavy neutrino masses are roughly equal, or
(ii) they are different.
\begin{itemize}

\item If all masses are equal, then we obtain
\beq
M > 2 \times 10^4 \sum_i{}^{\prime\prime} \left(U_{ei}\right)^2
{}~{\hbox{TeV}}~.
\label{equalmasslimit}
\eeq
In this case, if $\sum_i{}^{\prime\prime} \left(U_{ei}\right)^2$ is small,
$M$ will be as well. Note that, since the mixing angles may be complex, it
is possible that each of the individual $(U_{ei})^2$'s is large (up to the
constraint of Eq.~\ref{mixinglimit}), but that their sum is small.

\item In the second scenario involving quite different neutrino masses,
there can again be cancellations among different terms. This requires
either that the heavier neutrinos have larger mixings with the $\nu_e$ than
the lighter ones, or that there be a large number of heavy neutrinos. For
example, just to give a feel for the numbers, the contribution of a 1 TeV
neutrino with a mixing of $U^2 = 5 \times 10^{-3}$ can be cancelled by (a)
a 100 GeV neutrino with a mixing $U^2 = -5 \times 10^{-4}$ or (b) ten 10
TeV neutrinos with mixings of $U^2 = -5 \times 10^{-3}$. There are many
other possibilities, of course, but these illustrate roughly what is
required for cancellation.

\end{itemize}
We will return to these when discussing $\eeWW$.

\subsection{$\eeWW$ at the NLC}

The constraints from $\dbeta$ give us one of two conditions, depending on
whether the new neutrinos are very light (Eq.~\ref{lightNlimit}: $M \ll 1$
GeV) or very heavy (Eq.~\ref{heavyNlimit}: $M \gg 1$ GeV), relative to the
energy scale of neutrinoless double beta decay. For the case of light
neutrinos, we can use Eq.~\ref{Xsectionunitarity} to calculate the cross
section for $\eeWW$, which is independent of $\sqrt{s}$. It is minuscule:
\beq
\sigma(\eeWW) = 1.3 \times 10^{-17}~fb~.
\label{tinyXsection}
\eeq
Such a signal is clearly unobservable at any future collider.

If no cancellations are allowed in Eq.~\ref{heavyNlimit}, then
$\dbeta$ constrains the neutrinos to be very massive
(Eq.~\ref{lowerNlimit}). For NLC's with centre-of-mass energies of
order 1 TeV, we have $M_i \gg \sqrt{s}$, and Eq.~\ref{XsectionheavyN}
can be used to calculate the cross section for $\eeWW$. Using the
limit in Eq.~\ref{heavyNlimit}, we find that, at a $\sqrt{s}=1$ TeV
NLC,
\beq
\sigma(\eeWW) < 2.5 \times 10^{-3}~fb~.
\label{heavyNsigma}
\eeq
The hoped-for luminosity at a $\sqrt{s}=1$ TeV NLC is 80 $fb^{-1}$.
Clearly the process $\eeWW$ is unobservable at such a collider. (Since
the cross section grows like $s^2$, the 500 GeV NLC fares even worse.)

However, this does not cover all the possibilities. As discussed in the
previous subsection, the constraint from Eq.~\ref{heavyNlimit} can be
evaded if one allows cancellations among the various contributions. Thus we
must also consider neutrino masses considerably lighter than 100 TeV.
Nevertheless, as we discuss below, even for such masses the process $\eeWW$
is still unobservable at the NLC, except in the most contrived, fine-tuned
models.

We consider again the two scenarios for evading the constraint from
Eq.~\ref{heavyNlimit}: (i) roughly equal heavy neutrino masses, and
(ii) different heavy neutrino masses.
\begin{itemize}

\item In the scenario where all the neutrino masses are roughly equal,
there is an upper limit on the mixing as a function of the neutrino mass.
{}From Eq.~\ref{equalmasslimit} we have
\beq
\sum_i{}^{\prime\prime} \left(U_{ei}\right)^2 < 5 \times 10^{-5} \left(
{M \over {\hbox{1 TeV}}}\right).
\eeq
(Of course, even for super-heavy neutrinos, the mixing cannot be
larger than the phenomenological limit of Eq.~\ref{mixinglimit}.) It
is therefore possible to have neutrino masses lighter than 100 TeV,
but only at the expense of smaller mixing angles. This is the key
point. Even though the lighter neutrino masses soften, and even
remove, the $1/M^2$ suppression of Eq.~\ref{XsectionheavyN}, the
smaller mixing angles render the process $\eeWW$ unobservable. For $M$
in the range 500 GeV to 10 TeV, the cross section for $\eeWW$ is in
the range $O(10^{-4})$-$O(10^{-3})~fb$. In fact, the largest cross
section occurs for heavier neutrinos, $M\gsim 10$ TeV, where the
mixing angles are the largest. In this case, we simply reproduce the
cross section of Eq.~\ref{heavyNsigma}, which is, as we stated
previously, too small to be observed.

\item In the second scenario the cancellations occur between terms
involving neutrinos of quite different masses. This in itself is quite
contrived -- it requires a fair amount of fine tuning, since the masses and
mixing angles have to be carefully adjusted to have such a cancellation.
However, one has to go even further to obtain an observable cross section
for $\eeWW$.

One important observation is that a neutrino of mass $M < \sqrt{s}$ which
has a significant mixing with the $\nu_e$ would be first observed directly
at the NLC in the process $e^+e^- \to \nu_e N_l$ \cite{Abdel}. The decay
products of the $N_l$ would indicate that it is a Majorana neutrino. And
since such a neutrino would by itself violate the constraint from $\dbeta$
(Eq.~\ref{heavyNlimit}), one could deduce the presence of additional,
heavier Majorana neutrinos. Thus, if one has to add light ($M < \sqrt{s}$)
neutrinos in order to evade the constraints from Eq.~\ref{heavyNlimit} and
make the cross section for $\eeWW$ observable, then the measurement of the
process $\eeWW$ is not even necessary -- the neutrinos will be observed, or
their presence inferred, before $\eeWW$ is ever measured. A rather
amusing situation.

Suppose there were one heavy neutrino of mass $M \sim 1$ TeV, with a mixing
$U^2 = 5 \times 10^{-3}$. In this case the cross section for $\eeWW$ at a
1 TeV NLC is $\sigma \sim 10~fb$, which is easily observable. However, as
we have argued previously, if this is the only heavy neutrino, this set of
parameters is ruled out by the constraints from $\dbeta$. But if we add
other heavy neutrinos whose contributions conspire to evade the constraint
from $\dbeta$ (Eq.~\ref{heavyNlimit}), a neutrino with such a mass and
mixing could conceivably be allowed. One possibility is to add a lighter
neutrino $N_l$, say with mass $M=100$ GeV and a mixing $U^2 = -5 \times
10^{-4}$. However, as we have discussed above, such a light neutrino would
be first observed directly. Another possibility is to add ten neutrinos of
mass $M=10$ TeV and mixing $U^2 = -5 \times 10^{-3}$. This possibility is
clearly exceedingly baroque.

As a final example, if there were one heavy neutrino of $M \sim 1$ TeV,
with a mixing $U^2 = 5 \times 10^{-4}$, then the constraint from $\dbeta$
could be evaded through the addition of a single heavier neutrino of
$M=10$ TeV and mixing $U^2 = -5 \times 10^{-3}$. In this case, the cross
section for $\eeWW$ is $\sigma \simeq 0.04~fb$, which might be just
observable. Still, in addition to requiring the fine-tuned cancellation of
two terms, this scenario requires the heavier neutrino to have a {\it
larger} mixing angle than the lighter neutrino. This is rather unnatural,
and is not what happens in the quark sector.

\end{itemize}
Of course, there are many other ways of arranging the neutrino masses and
mixings in order to evade the low-energy constraint from $\dbeta$, and to
produce an observable cross section from $\eeWW$. However, the above
examples give a flavour of what is necessary -- one must construct
extremely contrived models in order to do this.

{}From here on, we assume that there are no fine-tuned cancellations,
and that the constraint in Eq.~\ref{lowerNlimit} holds for all
neutrinos. Furthermore, when we present our results for the cross
section for $\eeWW$ (and the other processes in the subsequent
sections), we assume that it is dominated by the exchange of a single
neutrino. (Of course, additional, heavier neutrinos must be present to
satisfy the bound from unitarity.) Even if one assumes that more than
one neutrino contributes to $\eeWW$, this will not change the cross
section significantly, since the mixing angles of all the neutrinos
must be correspondingly reduced in order to satisfy the constraint in
Eq.~\ref{heavyNlimit}.

\begin{figure*}[hbtp]
\vspace*{-5.5cm}
\begin{center}
\mbox{\epsfxsize=14cm\epsfysize=14.cm\epsffile{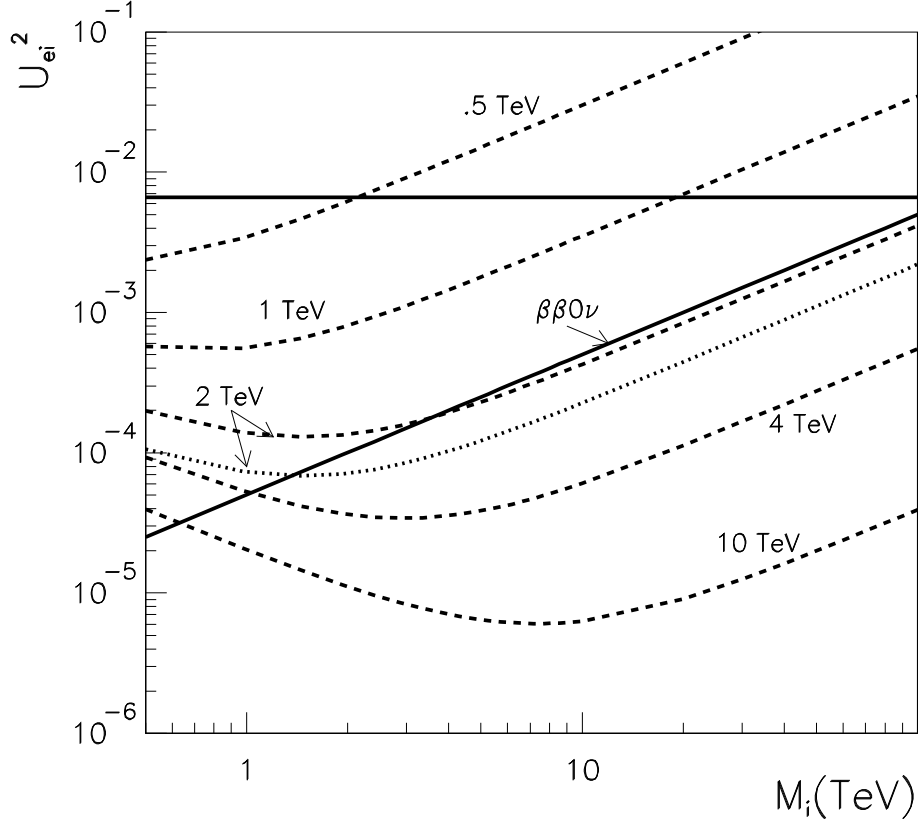}}
\vspace*{-1.cm}
\caption{\label{eeWWXsection}{\em Discovery limit for $\eeWW$ at the NLC
as a function of $M_i$ and $(U_{ei})^2$ for $\protect\sqrt{s}=500$ GeV, 1
TeV, 2 TeV, 4 TeV and 10 TeV (dashed lines). We assume unpolarized $e^-$
beams and a luminosity of $80 (\protect\sqrt{s}/{\hbox{(1
TeV)}})^2~fb^{-1}$. For $\protect\sqrt{s}=2$ TeV, the limit assuming
polarized $e^-$ beams is also shown (dotted line). In all cases, the
parameter space above the line corresponds to observable events. We also
superimpose the experimental limit from $\dbeta$ (diagonal solid line), as
well as the limit on $(U_{ei})^2$ (horizontal solid line). Here, the
parameter space above the line is ruled out.
}}
\end{center}
\end{figure*}

In Fig.~\ref{eeWWXsection} we present the discovery limit for $\eeWW$
at the NLC for several centre-of-mass energies as a function of $M_i$
and $(U_{ei})^2$. We demand 10 events for discovery, and assume
unpolarized $e^-$ beams and a luminosity of $80 (\sqrt{s}/{\hbox{(1
TeV)}})^2~fb^{-1}$. We present the discovery curves for $\sqrt{s}=
500$ GeV, 1 TeV, 2 TeV, 4 TeV and 10 TeV. We also superimpose the
phenomenological limit on $(U_{ei})^2$, as well as the constraint from
$\dbeta$. Note that we have not included efficiencies for the
detection of the $W$'s, nor have we included any backgrounds. Our
discovery limits are therefore quite conservative.

As is clear from this figure, for $\sqrt{s}=500$ GeV and 1 TeV, the values
of $M_i$ and $(U_{ei})^2$ which produce an observable $\eeWW$ cross section
are already ruled out by neutrinoless double beta decay. For $\sqrt{s}=2$
TeV, the discovery limit and the limit from $\dbeta$ are roughly equal.
Note, however, that if polarized $e^-$ beams are used, the 2 TeV NLC opens
a very small region of parameter space, and hence does slightly better than
$\dbeta$. On the other hand, by the time such a collider is built the
$\dbeta$ limits will probably have become more stringent, so the prospects
for a 2 TeV NLC to improve upon neutrinoless double beta decay are marginal
at best. Finally, for 4 TeV and 10 TeV NLC's, there exists a sizeable
region of $M_i$-$(U_{ei})^2$ parameter space, not ruled out by $\dbeta$,
which produces an observable signal for $\eeWW$.

\section{Other $\Delta L =2$ Processes at the NLC}

In the last section, we saw that the constraints from neutrinoless
double beta decay are so stringent that an NLC of at least $\sqrt{s} =
2$ TeV is required to be able to observe the process $\eeWW$. However,
$\dbeta$ constrains only the $\nu_e$ -- it says nothing about the
$\nu_\mu$ or the $\nu_\tau$. It therefore seems reasonable to ask
about the possibilities for observing other $\Delta L=2$ processes at
the NLC, specifically those involving a Majorana $\nu_\mu$ or
$\nu_\tau$. We address this issue in this section.

If the $\nu_\ell$ ($\ell = \mu,\tau$) is Majorana, it will mediate
processes such as $\llWW$. This is exactly like the $\nu_e$, except
that there are no constraints from $\dbeta$. On the other hand, there
is a major disadvantage -- the NLC involves $e^+$/$e^-$ beams, not
$\ell^+$/$\ell^-$. Thus, $\llWW$ cannot be observed directly as a $2
\to 2$ process at the NLC, unlike $\eeWW$. However, it does appear as
a subprocess in a number of $2\to 4$ processes involving the various
modes of the NLC. Specifically, if the $\nu_\ell$ is Majorana, it will
mediate $\ggllww$, $\egnllw$ and $\eennll$. (This is similar to the
analysis of Ref.~\cite{DKR}, where the process $pp\to (jet)_1 (jet)_2
e^+ e^+$ was considered.) We discuss these possibilities in turn in
the subsections which follow. In principle, the $e^+ e^-$ option of
the NLC can also be used: $e^+ e^- \to e^+ \nu_e \ell^- \ell^- W^+$.
However, since this is a $2\to 5$ process, it will be smaller than the
others, so we do not consider it further.

\subsection{Neutrino Masses and Mixing}

The limits on the masses of the $\nu_\mu$ and $\nu_\tau$ are \cite{PDG}
\beqn
m_{\nu_\mu} & < & 0.27~{\hbox{MeV}}, \cr
m_{\nu_\tau} & < & 31~{\hbox{MeV}}.
\eeqn
Suppose that the masses of the neutrinos are given by their upper
limits. If $\nu_\mu$ and $\nu_\tau$ are Majorana, but do not mix with
heavy neutrinos, then the cross section for $\llWW$ is still
unobservable -- from Eq.~\ref{tinyXsection} it is at most
$O(10^{-3})~fb$. Thus, in order to observe $\Delta L = 2$ processes
involving the $\nu_\mu$ or $\nu_\tau$, these neutrinos must mix with
heavy Majorana neutrinos, just as was the case for the $\nu_e$.

The limits on the mixing of the $\nu_\mu$ and $\nu_\tau$ are \cite{Nardi}
\beqn
\sum_{i\ne 1} |U_{\mu i}|^2 & < & 6.0 \times 10^{-3} ~~(90\%~c.l.)~, \cr
\sum_{i\ne 1} |U_{\tau i}|^2 & < & 1.8 \times 10^{-2} ~~(90\%~c.l.)~.
\label{nulmixing}
\eeqn
As with the $\nu_e$, these conservative limits are for the case where
the other fermions also mix with new particles. If one assumes that
only the neutrinos mix, then the limits improve to $1.8 \times
10^{-3}$ and $9.6 \times 10^{-3}$ for $\sum_{i\ne 1} |U_{\mu i}|^2$
and $\sum_{i\ne 1} |U_{\tau i}|^2$, respectively. In our analyses, we
will use the conservative limits above.

\subsection{$\ggllww$}

A large number of Feynman diagrams contribute to $\ggllww$. However,
it can be argued that a single one dominates. First, the diagrams can
be separated into two categories: ``fusion'' and ``bremmstrahlung.''
In the fusion diagrams, each photon splits into a real and a
quasi-real (\ie\ almost on-shell) particle. The two quasi-real
particles then interact, creating an internal $2\to 2$ process. In
bremmstrahlung diagrams, the two photons interact in a $2\to 2$
process, followed by the radiation of particles from one of the final
lines. The fusion diagrams are clearly much larger than the
bremmstrahlung diagrams, since they involve the propagators of almost
on-shell particles.

\begin{figure*}[hbt]
 \vspace*{-3.cm}
\begin{center}
 \mbox{\epsfxsize=17cm\epsfysize=20.cm\epsffile{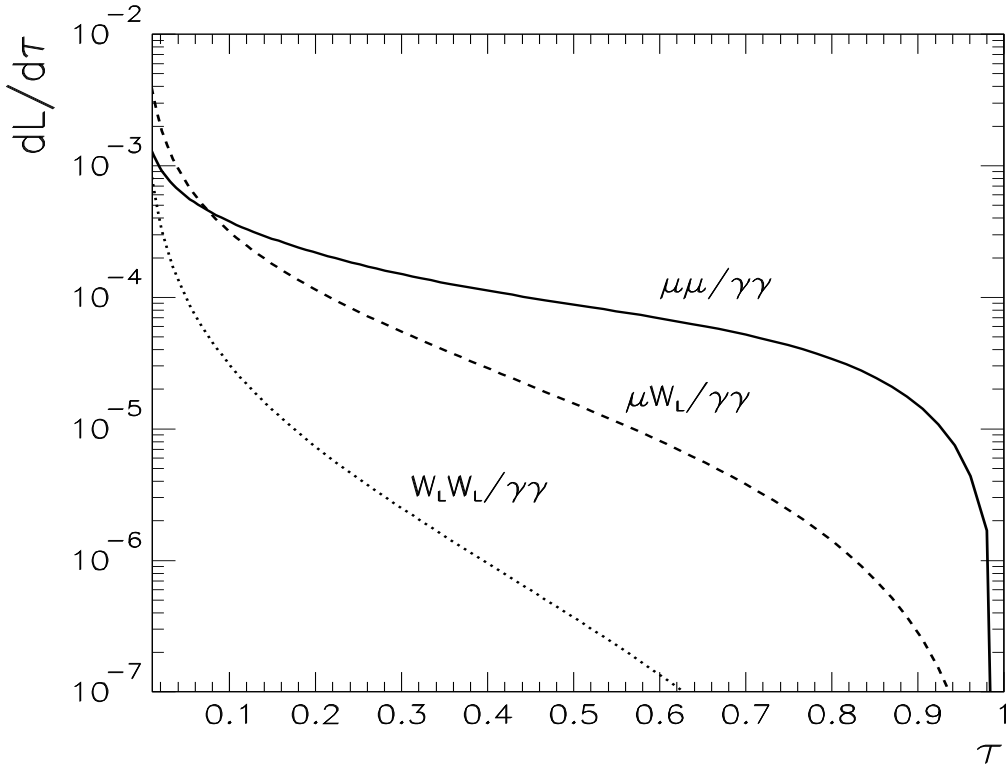}}
 \vspace*{-5.cm}
\caption{\label{strfuns}{\em The luminosity spectra for $\mu\mu$,
$\mu W_{long}$ and $W_{long}W_{long}$ in $\gamma\gamma$.}}
\end{center}
\end{figure*}

\begin{figure*}[hbt]
 \hspace*{2.cm}
 \vspace*{2.cm}
 \mbox{\epsfxsize=18cm\epsfysize=20.cm\epsffile{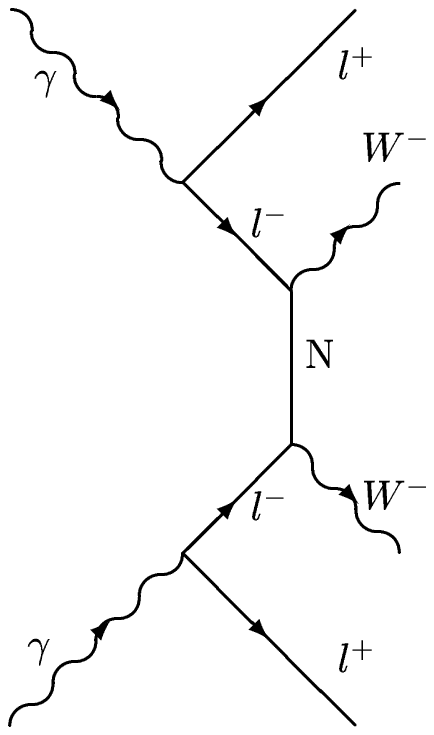}}
 \vspace*{-15.5cm}
\caption{\label{ggllwwdiag}{\em The dominant diagram in $\ggllww$.}}
\end{figure*}

There are 3 fusion diagrams, involving the internal $2\to 2$
subprocesses $\llWW$, $\ell^- W^+ \to \ell^+ W^-$ and $W^+ W^+ \to
\ell^+ \ell^+$. We remind the reader that it is primarily $W_{long}$,
the longitudinal component of the $W$, which is involved in the
subprocesses. In order to compare the sizes of these 3 fusion
diagrams, it is not necessary to calculate the entire $2\to 4$ process
-- one can simply convolute the internal $2\to 2$ process with the
structure functions of the $\ell$ and/or $W_{long}$ in the photon.
Thus, a comparison of the luminosity spectrum for $\ell\ell$, $\ell
W_{long}$ and $W_{long} W_{long}$ in $\gamma\gamma$ will suffice to
tell us which, if any, of the 3 fusion diagrams dominates.  In
Fig.~\ref{strfuns} we show the luminosity for $\ell=\mu$ and
$W_{long}$ as a function of the energy fraction
($\tau=\hat{s}/s_{\gamma\gamma}$) of the photons carried by the
quasi-real particles, $\mu$ or $W_{long}$. The luminosity is defined as
\beq
\frac{dL}{d\tau}=N \int dx f_{i/\gamma}(x,Q^2) f_{j/\gamma}(\tau/x,Q^2)~,
\eeq
where $N=1$ ($N=2$) if $i=j$ ($i\neq j$) and $i,j=\mu$ or $W_{long}$.
$Q^2$ is a typical scale for the subprocess. Here we take
$Q^2=s_{\gamma\gamma}/4$. The structure functions for the leptons are
taken from Ref.~\cite{eea} and those for the longitudinal $W$ were
given in Ref.~\cite{paris}{\footnote{The structure function describing
the $W_{long}$ content in the photon consists of two parts -- one where
the spectator W is transverse and the other where it is longitudinal.
It has been found that the former is much larger \cite{paris} and
shows scaling behaviour. Our numbers are based only on this
component.}}. It is clear that there is very little $W_{long}$ in the
photon, since over most of the energy range, and especially in the
high-energy region which gives the main contribution to the process
under study, $W_{long} W_{long} \ll \ell W_{long} \ll \ell\ell$. Thus,
the dominant diagram is the one in which the two quasi-real particles
are $\ell$ and the internal $2\to 2$ subprocess is $\llWW$. This is
shown in Fig.~\ref{ggllwwdiag}.

\begin{figure*}[hbt]
 \vspace*{-1.5cm}
\begin{center}
 \mbox{\epsfxsize=15cm\epsfysize=15.cm\epsffile{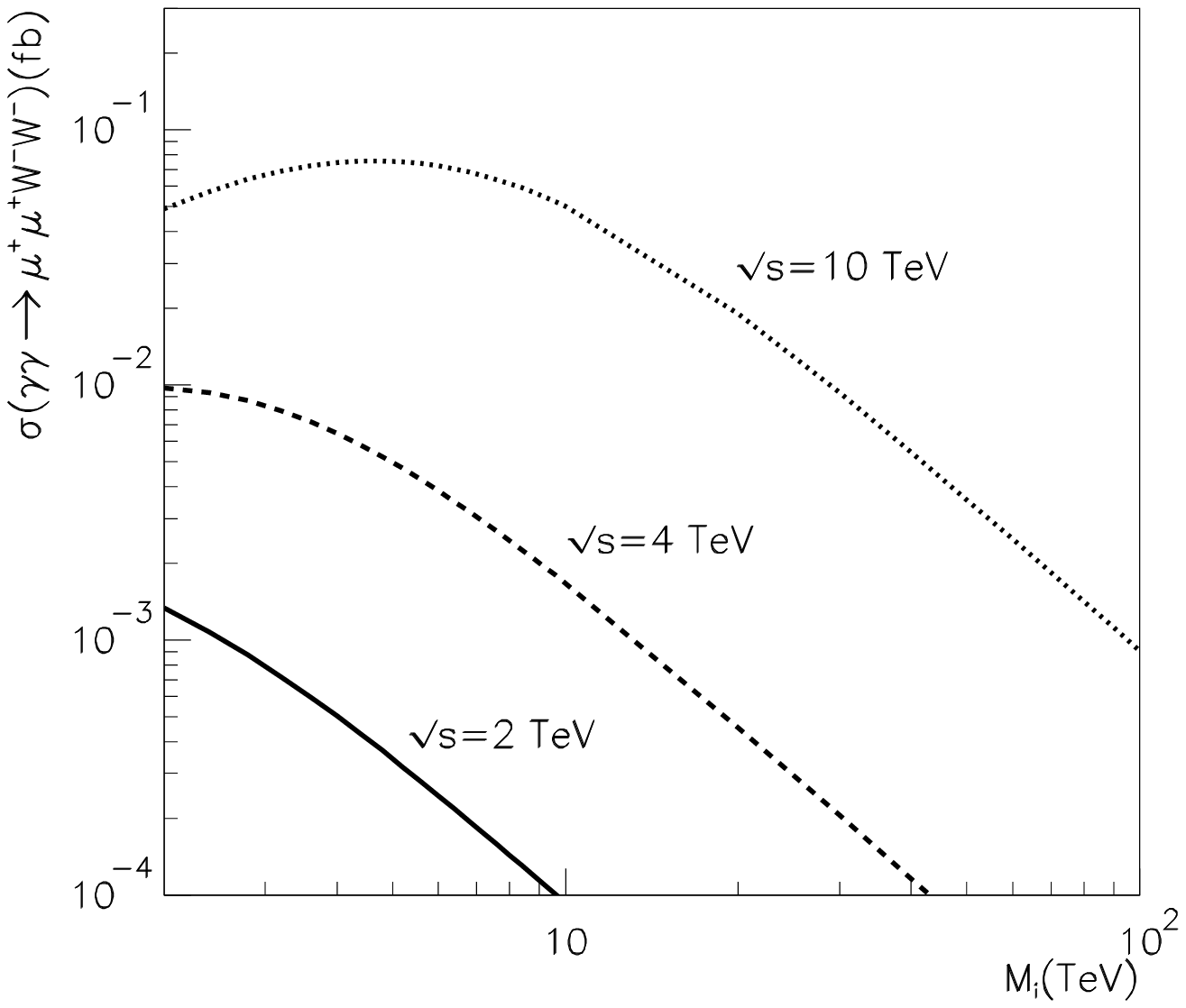}}
 \vspace*{-1.cm}
\caption{\label{ggmmwwXsection}{\em Cross section for $\ggmmww$ at the
NLC assuming $(U_{\mu i})^2 = 6.0
  \times 10^{-3}$ for $\protect\sqrt{s}=2$ TeV (solid line), 4 TeV
  (dashed line) and 10 TeV (dotted line). }}
\end{center}
\end{figure*}

In Fig.~\ref{ggmmwwXsection} we present the cross section for the process
$\ggmmww$ as a function of the neutrino mass $M_i$ for three centre-of-mass
energies: 2 TeV, 4 TeV and 10 TeV. We take $(U_{\mu i})^2=6.0 \times
10^{-3}$. Note that, in all cases, if $M_i < \sqrt{s}$, the new neutrino is
far more likely to be first discovered via single production in $e^+ e^-
\to \nu_\mu N_i$ than in $\ggmmww$ (see the discussion in Sec.~2.5). Thus,
although we present the cross section for a large range of neutrino masses,
we really should consider only $M_i > \sqrt{s}$. Assuming a luminosity of
$80 (\sqrt{s}/{\hbox{(1 TeV)}})^2~fb^{-1}$, we see that this process is
unobservable at $\sqrt{s} = 2$ TeV, regardless of the neutrino mass. And a
signal of 10 events can be observed at $\sqrt{s}=4$ TeV only for $M_i\lsim
3$ TeV. One has to go to higher energies to be able to observe $\ggmmww$
for $M_i > \sqrt{s}$: for example, for $\sqrt{s}=10$ TeV, the process is
observable for $M_i\lsim 90$ TeV. Of course, for the higher-energy NLC's,
the luminosity assumed is considerable -- the reality could be quite
different. But if the luminosity scales as we have assumed, and if the
neutrino mixing is as large as we have taken it to be, the $\Delta L=2$
process $\ggmmww$ can be observed at an NLC with a centre-of-mass energy
above 4 TeV.

We must also stress again that we have not considered here any
backgrounds and have only looked for processes with a few event
signals. A more careful analysis would include backgrounds from
standard processes without lepton number violation, such as
$\gamma\gamma \rightarrow \mu^+\mu^-W^+W^-$.
In addition, we have not folded in the photon energy
spectrum due to the backscattering of laser light off the $e^+$/$e^-$
beams. Since the backscattered photons are not monochromatic, the
inclusion of this spectrum would somewhat reduce the cross sections in
our figures.

There are certain numerical differences for the process $\ggttww$.
Although the mixing can be three times as large (see
Eq.~\ref{nulmixing}), there is also a suppression $(\ln[s/4 m_\tau^2]
/ \ln[s/4 m_\mu^2])^2$ due to the larger $\tau$ mass. Putting the
factors together, we estimate that the cross section for $\ggttww$ can
be roughly 4 times larger than that for $\ggmmww$. However, when one
folds in the much smaller efficiencies for detecting $\tau$'s, not to
speak of the increased backgrounds, it is more promising to search for
$\Delta L=2$ lepton number violation through $\ggmmww$.

\subsection{$\egnllw$}

\begin{figure*}[hbt]
 \hspace*{2cm}
 \vspace*{.3cm}
 \mbox{\epsfxsize=18cm\epsfysize=20.cm\epsffile{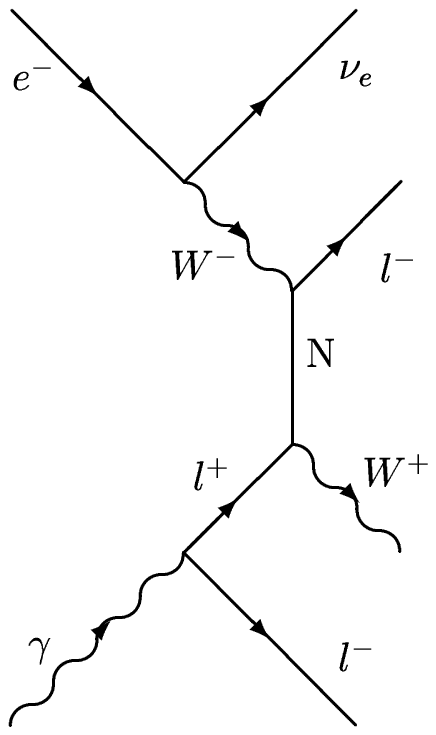}}
 \vspace*{-14cm}
\caption{\label{egnllwdiag}{\em The dominant diagram in $\egnllw$. }}
\end{figure*}

The process $\egnllw$ also involves a large number of Feynman
diagrams. However, just as was the case for $\ggllww$, there is a
single diagram which dominates, shown in Fig.~\ref{egnllwdiag}. (The
argument leading to this is essentially the same as for $\ggllww$.)
On the other hand, in contrast to $\ggllww$, note that this diagram
involves an internal $W_{long}$. Just like the photon, there is
relatively little $W_{long}$ in the electron (the dominant term in the
two sets of structure functions is the same up to a factor of $4
\sin^2\theta_w \approx 1$ \cite{paris}). We therefore expect the cross
section for $\egnllw$ to be suppressed relative to that for $\ggllww$.

\begin{figure*}[bht]
 \vspace*{-1.5cm}
\begin{center}
 \mbox{\epsfxsize=15cm\epsfysize=15.cm\epsffile{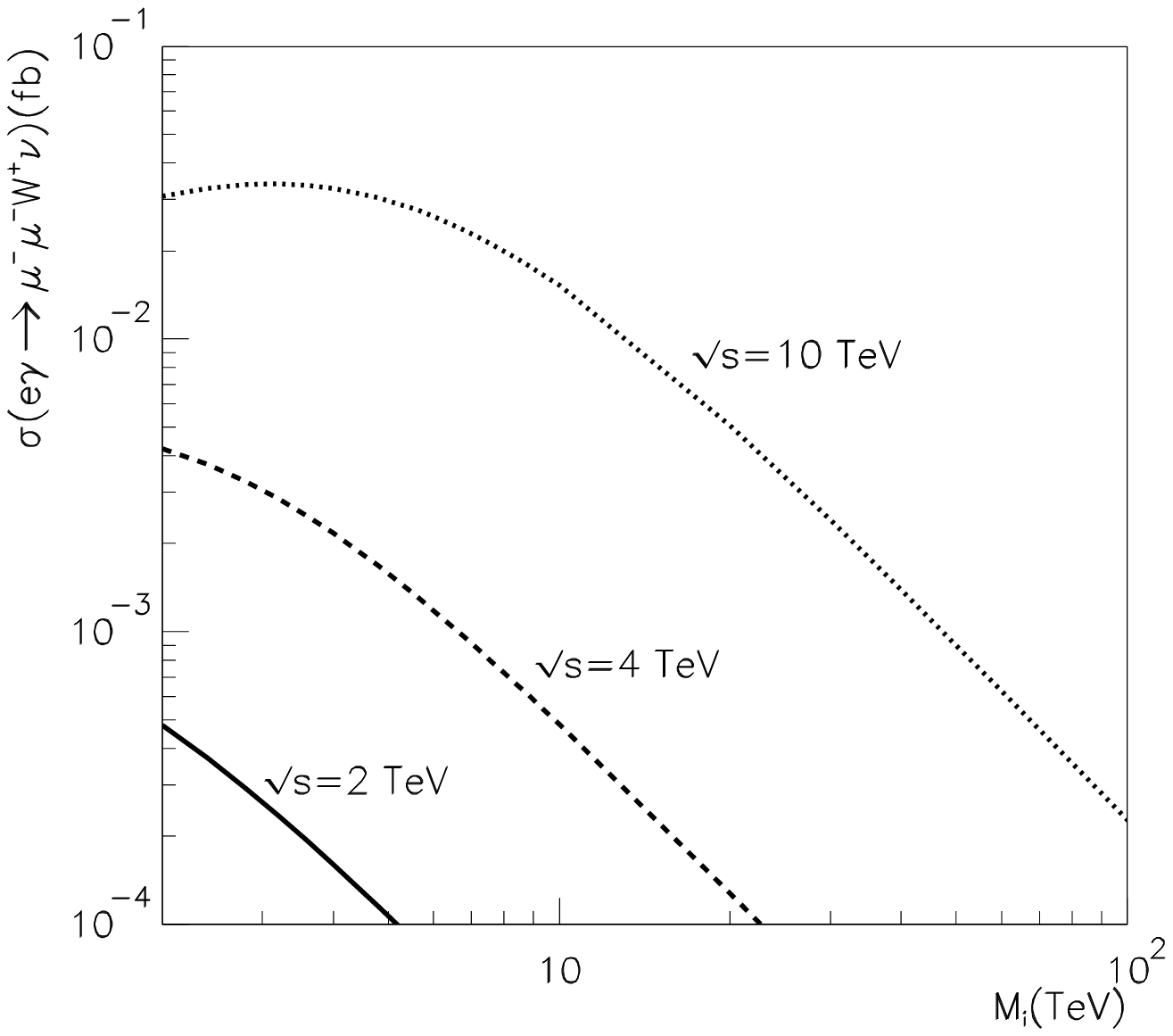}}
 \vspace*{-1.cm}
\caption{\label{egnmmwXsection}{\em Cross section for $\egnmmw$ at the NLC
assuming $(U_{\mu i})^2=6.0 \times 10^{-3}$ for $\protect\sqrt{s}=2$ TeV
(solid line), 4 TeV (dashed line) and 10 TeV (dotted line).}}
\end{center}
\end{figure*}

This is indeed the case. In Fig.~\ref{egnmmwXsection} we present the
cross section for $\egnmmw$ as a function of the neutrino mass $M_i$
for three centre-of-mass energies: 2 TeV, 4 TeV and 10 TeV. We again
take $(U_{\mu i})^2=6.0 \times 10^{-3}$. It is clear that the
situation is worse than for $\ggmmww$. Again assuming a luminosity of
$80 (\sqrt{s}/{\hbox{(1 TeV)}})^2~fb^{-1}$, we see that at
$\sqrt{s}=4$ TeV, this process is unobservable even for $M_i <
\sqrt{s}$. And at $\sqrt{s}=10$ TeV, the process is observable, but
the reach is reduced compared to $\ggmmww$ -- a signal of 10 events
requires $M_i \lsim 40$ TeV.

As far as $\egnttw$ is concerned, although the cross section may be
somewhat larger than that of $\egnmmw$, the process suffers from the
same problems as $\ggttww$: larger backgrounds and worse detection
efficiencies. When all is folded together, $\egnttw$ is probably worse
than $\egnmmw$.

\subsection{$\eennll$}

\begin{figure*}[hbt]
 \hspace*{2.cm}
 \vspace*{1cm}
 \mbox{\epsfxsize=18cm\epsfysize=20.cm\epsffile{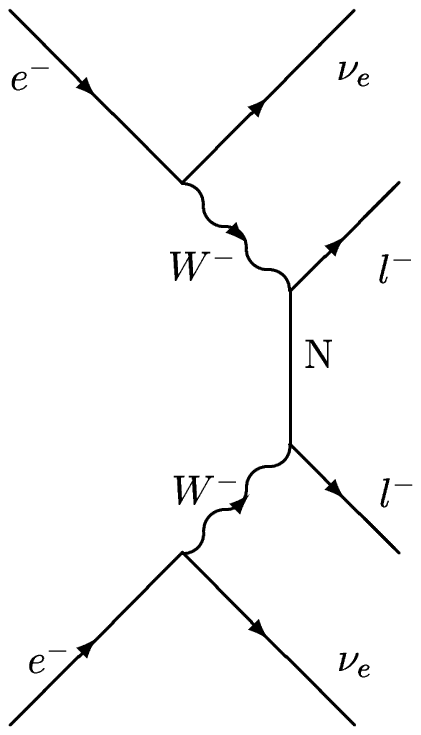}}
 \vspace*{-14.cm}
\caption{\label{eennlldiag}{\em The dominant diagram in $\eennll$.}}
\end{figure*}

Should the $e^- e^-$ option of the NLC be available, not only can it
be used to search for a Majorana $\nu_e$ via $\eeWW$, as discussed in
Sec.~2, but a Majorana $\nu_\ell$ ($\ell=\mu,\tau$) can in principle
be detected through the process $\eennll$. The diagram is shown in
Fig.~\ref{eennlldiag}. However, note that this diagram involves two
internal $W_{long}$'s. Therefore the cross section for this process
will be suppressed relative to that for $\egnllw$, and very suppressed
relative to $\ggllww$. It is clear, therefore, that $\eennll$ is far
from the optimal way to search for $\Delta L=2$ processes involving a
Majorana $\nu_\mu$ or $\nu_\tau$, and we do not consider it further.

\section{Detecting a Majorana $\nu_e$ Without an $e^-e^-$ Collider}

\begin{figure*}[hbt]
 \hspace*{1.5cm}
 \vspace*{.3cm}
 \mbox{\epsfxsize=18cm\epsfysize=20.cm\epsffile{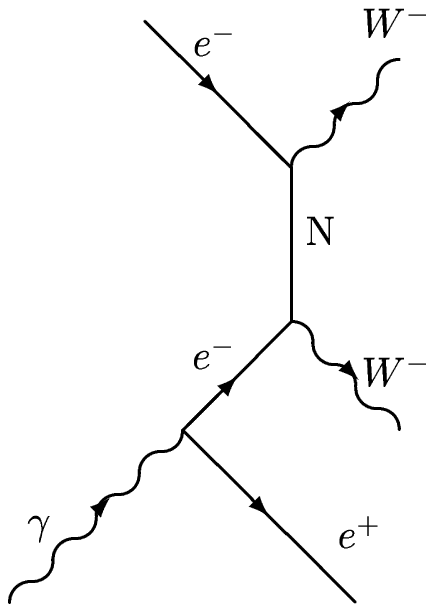}}
 \vspace*{-14.5cm}
\caption{\label{egewwdiag}{\em The dominant diagram in $\egeww$.}}
\end{figure*}

For various reasons, it is conceivable that, even if an NLC is built, the
$e^- e^-$ option may never be used. In the absence of an $e^-e^-$ collider,
what are the prospects for detecting a Majorana $\nu_e$ through $\Delta
L=2$ processes similar to those discussed to this point? The only real
possibility is the $2\to 3$ process $\egeww$.
{\footnote{The $\Delta L=2$ process $\ggeeww$
could also occur, with a cross-section slightly
larger than the one presented for $\ggmmww$ due to the
larger electron content of the photon. However,
with the constraint from $\dbeta$, a few events are expected
only for very heavy neutrinos.}}
 The dominant diagram for this
process is shown in Fig.~\ref{egewwdiag}.

\begin{figure*}[hbtp]
 \vspace*{-6.cm}
\begin{center}
 \mbox{\epsfxsize=15cm\epsfysize=15.cm\epsffile{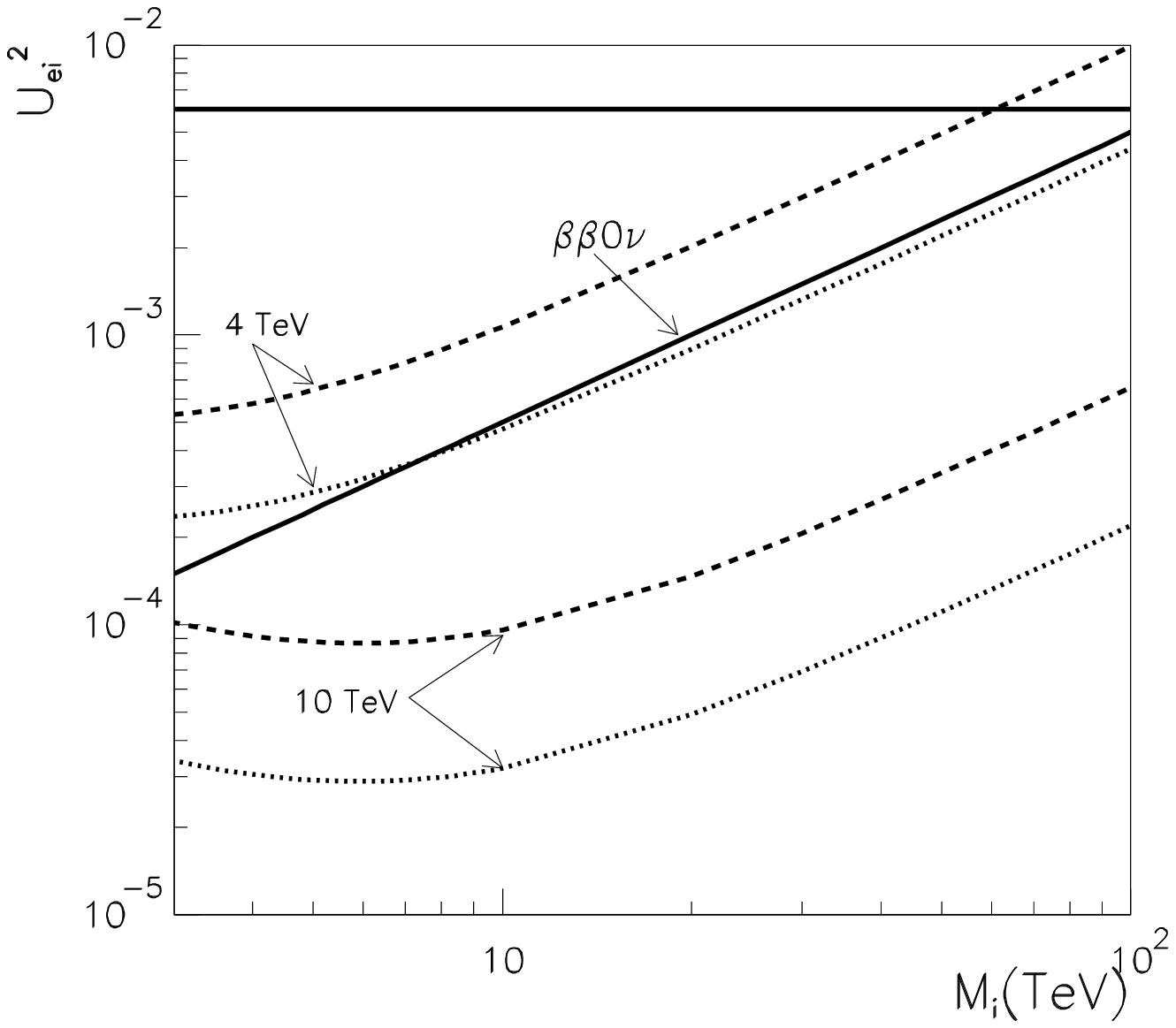}}
 \vspace*{-1.cm}
\caption{\label{egewwXsection}{\em Discovery limit for $\egeww$ at the NLC
as a function of  $M_i$ and $(U_{ei})^2$. We assume a luminosity of $80
(\protect\sqrt{s}/{\hbox{(1 TeV)}})^2~fb^{-1}$.  The dash-dot (dotted) line
corresponds to an unpolarized (polarized) $e^-$ beam, and we require 25
(10) events for discovery. In all cases, the parameter space above the line
corresponds to observable events. We also superimpose the experimental
limit from $\dbeta$ (diagonal solid line), as well as the limit on
$(U_{ei})^2$ (horizontal solid line). Here, the parameter space above the
line is ruled out.}}
\end{center}
\end{figure*}

In Fig.~\ref{egewwXsection} we present the discovery limit for
$\egeww$ at the NLC for $\sqrt{s}=4$ TeV and 10 TeV as a function of
$M_i$ and $(U_{ei})^2$. We consider two scenarios. In the optimistic
(conservative) scenario, the $e^-$ is polarized (unpolarized), and we
demand 10 (25) events for discovery. As is clear from the figure, for
$\sqrt{s}=4$ TeV, even in the optimistic scenario the values of $M_i$
and $(U_{ei})^2$ which produce an observable $\egeww$ cross section
are already ruled out by neutrinoless double beta decay. However, for
$\sqrt{s}=10$ TeV, there exists a sizeable allowed region of
$M_i$-$(U_{ei})^2$ parameter space which produces an observable signal
for $\egeww$. Therefore, even if the $e^-e^-$ option of the NLC is
never used, it will be possible to detect a Majorana $\nu_e$. However,
one must go to extremely high energies and luminosities.

\section{Conclusions}

We have critically reexamined the prospects for the observation of
$\eeWW$ at a high-energy $e^-e^-$ collider. This process is
essentially the inverse of neutrinoless double beta decay ($\dbeta$).
Once the constraints from $\dbeta$ are taken into account, we have
found that $\eeWW$ is unobservable at an NLC of $\sqrt{s} <2$ TeV. It
is possible to evade the constraints, but this requires models which
are extremely contrived and fine-tuned. A $\sqrt{s}=2$ TeV NLC
essentially reproduces the limits from $\dbeta$, and for $\sqrt{s}>2$
TeV, there is a sizeable region of parameter space, not ruled out by
$\dbeta$, which produces an observable signal for $\eeWW$.

The constraints from $\dbeta$ apply only to Majorana neutrinos which
mix with the $\nu_e$. $\Delta L=2$ processes in the $\mu$ or $\tau$
sectors are unconstrained by $\dbeta$. We have therefore also
considered other $\Delta L=2$ processes at the NLC, involving $\mu$-
and $\tau$-lepton-number violation. The process $\ggllww$
($\ell=\mu,\tau$) can be observed for $\sqrt{s}>4$ TeV, while the
observation of $\egnllw$ requires $\sqrt{s} \sim 10$ TeV.

Finally, we have examined the possibilities for the observation of
$\Delta L=2$ $e$-lepton-number violation in the absence of an $e^-e^-$
collider. The most promising process is $\egeww$. Taking into account
the constraints from $\dbeta$, we have found that its observation
requires $\sqrt{s} \sim 10$ TeV.

\bigskip
\noindent
Note added: While writing up this paper, we received Ref.~\cite{GZ},
which also discusses $\eeWW$. These authors arrive at the conclusion
that this process is observable at a 1 TeV NLC. However, like
Ref.~\cite{HM}, they have not included the constraints from $\dbeta$.

\bigskip
\centerline{\bf ACKNOWLEDGMENTS}
\bigskip

We thank J. Cline for enlightening conversations and B. Kayser for
helpful communications. D. London and H. Nadeau are grateful for the
hospitality of ENSLAPP, where most of this work was done. This work
was supported in part by the NSERC of Canada and les Fonds FCAR du
Qu\'ebec.

\section{Appendix}

The helicity amplitudes for $\eeWW$ can be written in a very compact
way. The only non-vanishing helicity amplitudes are those involving
left-handed electrons, \ie\ we have $2 \lambda_e=2\lambda_e'=-1$
where $ \lambda_e$ is the helicity of the electron. The helicities of
the $W^-$ are denoted by $h_i$, with $h_i=0,\tau$, $\tau=\pm 1$ and
$0$ is the longitudinal contribution. $\beta=\sqrt{1-4M_W^2/s}$ will
denote the velocity of the $W$ in the centre-of-mass system.

The amplitude is given by
\beqn
{{\cal M}}^{\lambda,\lambda'}_{h_1,h_2}=\frac{\delta_{2\lambda,-1}\;
\delta_{2\lambda',-1}}{2\sqrt{s}}
\left(\frac{g}{\sqrt{2}}\right)^2 \sum_j M_j \; U^2_{ej} \;
A_{h_1,h_2}^{(j)} ~,
\eeqn
where
\beqn
A_{00}^{(j)}&=&-\frac{2 s}{M_W^2} \left( \frac{t}{t-M_j^2} +
\frac{u}{u-M_j^2} \right)  \nonumber \\
A_{\tau,\tau'}^{(j)}&=& 2 \delta_{\tau,\tau'} \left\{\frac{s}{t-M_j^2}
\left( 1- \frac{\tau}{\beta}
\frac{t-u}{s}\right) + t \leftrightarrow u \right\}  \nonumber \\
A_{0,\tau}^{(j)}&=&\sqrt{\frac{s}{M_W^2}} \left(\frac{u-t}{\beta}
\sqrt{\frac{2}{ut -M_W^4}} \right)
\frac{s^2}{(t-M_j^2) (u-M_j^2)} \nonumber \\
&&\left\{ \frac{ut-M_W^4}{s^2} - \frac{\tau}{\beta} \left( \frac{(t+M_W^2)
(u+M_W^2)}{s^2}
+\frac{M_W^2}{s} \frac{(t-u)^2}{s^2} \right) \right\}~.
\eeqn


\end{document}